\documentclass{andp2012}% no class options needed by now
\usepackage[english]{babel}
\usepackage{setspace}
\usepackage{graphicx}
\usepackage{amssymb,amsmath}
\usepackage{subfigure}
\usepackage{graphicx}
\usepackage{bbm}
\usepackage{stmaryrd}
\usepackage{xcolor}
\usepackage{rotating}
\newcommand\ii{\mathrm{i}}
\newcommand\dd{d\!}
\newcommand\ee{\mathrm{e}}

\keywords{Hubbard model, slave boson, radial gauge}
\title{Combining complex and radial slave boson fields within the
  Kotliar-Ruckenstein representation of correlated impurities} 
\author[V. H. Dao]{Vu Hung Dao\inst{1}}
\author[R. Fr\'esard]{Raymond Fr\'esard\inst{1,}\footnote{Corresponding
    author\quad 
			E-mail:~\textsf{Raymond.Fresard@ensicaen.fr},
			Phone: +33\,231\,45\,26\,09,
			Fax: +33\,231\,95\,16\,00}}
\address[1]{Normandie Universit\'e, ENSICAEN, UNICAEN, CNRS, CRISMAT, 14000
  Caen, FRANCE} 
\shortauthors{V. H. Dao et al.}

\begin{abstract}
The gauge symmetry group of any slave boson representation allows to gauge 
away the phase of bosonic fields. One benefit of this radial field formulation 
is the elimination of spurious Bose condensations when saddle-point 
approximation is performed. Within the Kotliar-Ruckenstein representation,
three of the four bosonic fields can be radial while the last one has to
remain complex. In this work, we present the procedure to carry out the
functional integration involving constrained fermionic fields, complex bosonic
fields, and radial bosonic fields. The correctness of the representation is
verified by exactly evaluating the partition function and the Green's
function of the Hubbard model in the atomic limit. 
\end{abstract}
\shortabstract
\begin{document}
\maketitle

\section{Introduction}

The challenges posed by strongly correlated electron systems have been tackled 
using slave-boson approaches in a number of ways. The most popular ones are
probably Barnes' \cite{Barnes76,Barnes77} and Kotliar and Ruckenstein's (KR)
\cite{KR} 
representations, as well as their multiband and rotation invariant
generalizations \cite{FK,LiWH,FW92,Kotliar-Georges,Piefke}. For 
instance, it has been proposed that strong correlation effects taking place
in the plates of a capacitor can enhance its capacitance \cite{Ste17}. 

Any slave boson representation possesses a gauge symmetry group. Regarding  
Barnes' representation to the infinite-$U$ single-impurity Anderson model, it 
involves one doublet of fermions, one slave boson, and one time-independent 
constraint. Its $U(1)$ gauge symmetry may be used to gauge away the phase of 
the slave boson at the price of introducing a time-dependent constraint 
field~\cite{RN83a,RN83b,NR87}. While the argument was initially put forward in
the continuum limit, a path integral representation on discrete time steps for
the 
remaining radial slave boson together with the time-dependent constraint has 
been proposed by one of us~\cite{RFTK01}. A toy model has been analyzed, with 
the result that the exact expectation value of the radial slave boson is 
generically finite~\cite{RFHOTK07}. Thereby a way to exactly handle functional 
integrals involving constrained fermions and radial slave bosons could be put 
forward. Furthermore, a saddle-point approximation involving the complex slave
boson field is intimately tied to a Bose condensation, that is generally viewed
as spurious. On the contrary, radial slave bosons do not Bose condense, as
they are deprived of a phase degree of freedom, and their saddle-point
amplitude approximates their -- generically finite -- exact expectation value. 

The KR representation~\cite{KR}, and related slave-boson
representations~\cite{LiWH,FW92}, have been set up for the Hubbard model,
allowing to obtain a whole range of valuable results. In particular, 
they have been used to describe anti-ferromagnetic~\cite{Lil90}, 
spiral~\cite{Fre91,Igo13,Fre92,Doll2}, and
striped~\cite{SeiSi,Fle01,Sei02,Rac06a,Rac06b} phases. Furthermore, the
competition 
between the latter two has been  
addressed as well~\cite{RaEPL}. Besides, it has been obtained that the 
spiral order continuously evolves to the ferromagnetic order in the
large $U$ regime ($U \gtrsim 60t$)~\cite{Doll2} so that it is unlikely to be
realized 
experimentally. Consistently, in the two-band model, ferromagnetism was
found as a possible groundstate only in the doped Mott insulating 
regime~\cite{Fre02}. Yet, adding a ferromagnetic exchange coupling was 
shown to bring the ferromagnetic instability line into the intermediate
coupling regime~\cite{lhoutellier15}. A similar effect has been obtained
with a sufficiently large next-nearest-neighbor hopping amplitude~\cite{FW98} or
going to the fcc lattice~\cite{Igo15}.

The KR representation again implies one 
doublet of fermions, but now subjected to four slave bosons through three 
constraints. The gauge symmetry group of the representation has been debated 
over the years~\cite{Ras88,Lil90,Lav90,Li94}, but a consensus that it reads 
$U(1)\times U(1) \times U(1)$ has been reached~\cite{Jol91,FW92,Kot92}. Hence, 
the phase of three of the four slave bosons can be gauged away, therefore 
leaving us with one single complex field. It can be the $d$-field which 
accounts for doubly occupied sites, or the $e$-field which accounts for empty 
sites. These fields are primarily associated with charge fluctuations, and one
might wonder about the consequences of this asymmetry on the charge 
fluctuation spectrum. It turns out that the latter is independent of the 
choice, at least when computed to one-loop order around the paramagnetic 
saddle-point~\cite{Dao17,Dao18}.

Explicitly taking the above gauge symmetry into account yields a new type of 
problem from the path integral point of view: how can one simultaneously
handle constrained fermionic fields, complex bosonic fields, and radial boson
fields? The purpose of this work is to show that the partition function and the 
Green's function of the isolated correlated impurity can be exactly 
calculated in this situation, in contrast to earlier works where all slave
boson fields were taken as radial fields~\cite{RFTK01,RFHOTK07,RFTK12}.

\section{Model}
We investigate an exactly soluble model, the single-site Hubbard Hamiltonian
\begin{equation}\label{eqh1}
{\mathcal H} 
= \epsilon  \sum_{\sigma} c^{\dagger}_{\sigma} c^{\phantom{\dagger}}_{\sigma}
+ U c^{\dagger}_{\uparrow} c^{\phantom{\dagger}}_{\uparrow}  
c^{\dagger}_{\downarrow} c^{\phantom{\dagger}}_{\downarrow}   .
\end{equation}
The KR representation of this model implies one doublet of fermions 
$\{ f_{\uparrow}, f_{\downarrow} \}$ and four slave bosons 
$\{e, p_{\uparrow}, p_{\downarrow},d\}$. 
The latter are tied to an empty site, single occupancy of the site 
with spin projection up or down, and double occupancy, respectively.
The redundant degrees of freedom are discarded provided the three constraints
\begin{eqnarray}\label{eqrf:q2_KR}
e^\dagger e + \sum_\sigma p^{\dagger}_{\sigma}p^{\phantom{\dagger}}_\sigma
+ d^\dagger d &=&1 \nonumber\\
f^{\dagger}_{\sigma} f^{\phantom{\dagger}}_\sigma
-p^{\dagger}_{\sigma}p^{\phantom{\dagger}}_\sigma 
- d^\dagger d &=& 0  \;\;\;\;\;\;\;\;\;\;(\sigma
=\uparrow,\downarrow)
\end{eqnarray}
are satisfied. In the functional integral formalism the latter are enforced 
by the Lagrange multipliers $\lambda$, resp. $\lambda_{\sigma}$. At this stage
it should be noted that the functional integral over the fermionic and bosonic
fields cannot be  
performed right away. Indeed, in order to ensure the convergence, $\lambda$ 
has to be continued into the complex plane as 
\begin{equation}
\tilde{\lambda} \equiv \lambda - \ii \lambda_0  
\end{equation}
with $\lambda_0 > 0$ (or $\lambda_0 + U >0$ if $U<0$) so that the integration 
contour is shifted into the lower half-plane \cite{Bickers86,Bickers87,RFTK01}.

Hence the Lagrangian
\begin{equation}\label{eqLtot}
{\mathcal{L}}(\tau) = {\mathcal{L}}_f(\tau) + {\mathcal{L}}_b(\tau) 
\end{equation}
consists of a fermionic contribution,
\begin{equation}\label{Lcont_f}
{\mathcal{L}}_f(\tau) = \sum_{\sigma}
f^{\dagger}_{\sigma}(\tau) \left(\partial_{\tau} + \epsilon  +
\ii \lambda_{\sigma}\right)f^{\phantom{\dagger}}_{\sigma}(\tau) ,
\end{equation}
and a bosonic one,
\begin{eqnarray}\label{Lcont_b}
{\mathcal{L}}_b(\tau) &=& - \ii\lambda  + e^{\dagger}(\tau) 
\left(\partial_{\tau} + \ii \lambda\right) e(\tau) 
\nonumber\\
&\quad& + \; d^{\dagger}(\tau) \left(\partial_{\tau} + U + \ii\lambda 
- \ii \lambda_{\uparrow} - \ii \lambda_{\downarrow}\right) d(\tau)  
\nonumber \\
&\quad& +\; \sum_{\sigma} p^{\dagger}_{\sigma}(\tau) \left(\partial_{\tau} 
+ \ii\lambda - \ii \lambda_{\sigma} \right) p^{\phantom{\dagger}}_{\sigma}(\tau) .
\end{eqnarray}
It entails the dynamics of the auxiliary fermionic and bosonic fields, 
together with the constraints~(\ref{eqrf:q2_KR}) specific to the Kotliar and 
Ruckenstein setup. 

In this representation, the physical electron creation and annihilation
operators read 
\begin{equation}\label{eq:KR}
  \begin{array}{l}
          c_{\sigma}^{\phantom{\dagger}} \leadsto \left( e^{\dagger} 
          p_{\sigma}^{\phantom{\dagger}} + p_{-\sigma}^{\dagger} d \right) 
          f_{\sigma}^{\phantom{\dagger}} 
          \\
          c_{\sigma}^{\dagger} \leadsto \left( p_{\sigma}^{\dagger} e + 
          d^{\dagger} p_{-\sigma}^{\phantom{\dagger}} \right) 
          f_{\sigma}^{\dagger} 
         \end{array}    
\end{equation}
This representation of the physical electron operators is invariant under the 
gauge transformations
\begin{equation}
  \left\{ \begin{array}{l}
          f_\sigma \longrightarrow \ee^{-\ii \chi_\sigma} 
          f_\sigma
          \\
          e_{\phantom{\sigma}} \longrightarrow \ee^{\ii \theta} e
          \\
          p_{\sigma} \longrightarrow \ee^{\ii \left(\chi_\sigma + \theta \right)}
          p_{\sigma}
          \\
          d_{\phantom{\sigma}} \longrightarrow \ee^{\ii \left(\chi_\uparrow + 
          \chi_\downarrow  + \theta\right)} d
         \end{array} 
   \right.      
\end{equation}
The gauge symmetry group is therefore $U(1) \times U(1) \times U(1)$.
The Lagrangian, Eq.~(\ref{eqLtot}), also possesses this symmetry. 
Expressing the bosonic fields in amplitude and phase variables as
\begin{eqnarray}
e(\tau) & = & \sqrt{R_e(\tau)}\, \ee^{\ii \theta(\tau)}\nonumber\\
p^{\phantom{\dagger}}_\sigma (\tau) & = & \sqrt{R_\sigma(\tau)}\,
\ee^{\ii (\chi_\sigma(\tau) + \theta(\tau))} 
\end{eqnarray}
allows to gauge away the phases of three of the four slave boson fields 
provided one introduces the three time-dependent Lagrange multipliers
\begin{eqnarray}
\alpha (\tau) & \equiv & \lambda + \partial_{\tau}
\theta(\tau) \nonumber\\
\beta_\sigma (\tau) & \equiv &  \lambda_{\sigma}
  - \partial_{\tau} \chi_{\sigma}(\tau) .
\end{eqnarray}
Here the radial slave boson fields are implemented in the continuum limit
following, e.~g., Ref.~\cite{RN83a,RN83b,NR87}.  

\section{Partition function}

\subsection{Cartesian slave boson representation}
First we evaluate the partition function with the Cartesian representation 
of the slave boson fields:
\begin{eqnarray}\label{Zcan}
{\mathcal Z} &=& \int_{-\pi/\beta}^{\pi/\beta} \frac{\beta \dd\lambda }{2\pi}  
\prod_{\sigma} \frac{\beta \dd\lambda_{\sigma}}{2\pi}
\int \prod_{\sigma} D[f^{\phantom{\dagger}}_{\sigma},f^{\dagger}_{\sigma}] 
\nonumber \\
&\quad& \times  \int D[e,e^{\dagger}]  D[d,d^{\dagger}] 
\prod_{\sigma} D[p^{\phantom{\dagger}}_{\sigma},p^{\dagger}_{\sigma}]
\nonumber \\
&\quad& \times \;\; \ee^{-\int_0^{\beta} d \tau {\mathcal{L}}(\tau)} .
\end{eqnarray}
Note that the constraints are constants of motion 
since they commute with the Hamiltonian, which does not hybridize the 
auxiliary boson states. They may hence be enforced by time-independent 
constraints. Therefore one can first sum over all states, and afterward 
restrict the partition function to the physical subspace by imposing the 
constraints. 

Integrating the auxiliary fermionic and bosonic fields yields
\begin{eqnarray}
&{\mathcal Z} &= \int_{-\pi/\beta}^{\pi/\beta} \frac{\beta \dd\lambda }{2\pi}  
\prod_{\sigma} \frac{\beta \dd\lambda_{\sigma}}{2\pi} \\
&\times& \frac{\ee^{\ii \beta \tilde{\lambda}} \prod\limits_{\sigma} 
\left( 1 + \ee^{-\beta(\epsilon + \ii \lambda_{\sigma})}\right)}
{\left( 1 - \ee^{-\ii \beta \tilde{\lambda}} \right) 
\left( 1 - \ee^{-\beta \left(U + \ii \left(\tilde{\lambda} 
- \lambda_{\uparrow} - \lambda_{\downarrow} \right) \right)} \right) 
\prod\limits_{\sigma} \left( 1 - \ee^{-\ii \beta \left(\tilde{\lambda} 
- \lambda_{\sigma} \right)} \right) } . \nonumber
\end{eqnarray}
At this point one may introduce new variables 
$x\equiv \ee^{-\ii \beta \lambda_{\uparrow}}$, 
$y \equiv \ee^{-\ii \beta \lambda_{\downarrow}}$, and 
$z \equiv \ee^{-\ii \beta \lambda}$, so that evaluating $\mathcal Z$ results 
in contour integrals along the complex unit circle $\mathbb{U}$ in the 
clockwise direction:
\begin{eqnarray}
&{\mathcal Z} &=  \,  \left( \frac{\ii }{2\pi} \right)^3  \oint_{\mathbb U} 
 \dd x \, \dd y \, \dd z \\ 
&\times& \frac{  x\, y\: \ee^{\beta \lambda_0} 
\left( 1 + x \ee^{-\beta \epsilon } \right) 
\left( 1 + y \ee^{-\beta \epsilon } \right) }
{ z^2 (1 - z\ee^{-\beta \lambda_0}) ( xy - z \ee^{-\beta (U+\lambda_0)})
( x - z \ee^{-\beta \lambda_0}) ( y - z \ee^{-\beta \lambda_0})} . \nonumber
\end{eqnarray}
One is then ready to evaluate the integral over $z$. Since $\lambda_0>0$ for
positive $U$ ($\lambda_0+U>0$ otherwise), all poles are located outside
$\mathbb{U}$ but the one at the origin. Hence, the 
$z$-contour integral only picks up the residue at $z=0$. Thus
\begin{eqnarray}
\label{EQ:zmathcal}
{\mathcal Z} & = & \left( \frac{\ii }{2\pi} \right)^2  \oint_{\mathbb U} 
 \dd x \, \dd y \frac{1}{xy} \left( 1 + \frac{1}{x} + \frac{1}{y} + 
 \frac{ \ee^{-\beta U}}{x y} \right)  \nonumber \\
 &\quad& \times\; \left(1 + x \ee^{-\beta \epsilon  } \right)
 \left( 1  + y \ee^{-\beta \epsilon } \right) , 
\end{eqnarray}
which is actually independent of the value of $\lambda_0$. Yet, we remind the
reader that it takes a positive value of $\lambda_0$ to obtain
Eq.~(\ref{EQ:zmathcal}). Finally the remaining integrations yield the expected
result 
\begin{equation}
{\mathcal Z} = 1 +  2 \ee^{-\beta \epsilon} + \ee^{-\beta ( 2 \epsilon + U ) } .
\end{equation}
This gives a strong indication that this representation is faithful.

\subsection{Radial slave boson representation}

For the exact evaluation of the functional integrals, the representation in 
the radial gauge has to be set on a discretized time mesh from the beginning. 
Moreover, the constraints now have to be satisfied at every time step. 
Extending the procedure introduced in Ref.~\cite{RFTK01} for Barnes' slave 
boson to the Kotliar and Ruckenstein representation one can cast the 
partition function
\begin{equation}\label{eqh2}
{\mathcal Z} = \lim_{N\rightarrow\infty} \lim_{\eta\rightarrow 0^{+}}
{\mathcal P} {Z}_f {Z}_d
\end{equation}
as a projection onto the physical subspace of the product
of the auxiliary fermions partition function with the $d$ boson  partition 
function. Here we introduced 
\begin{equation}\label{eqprojp}
{\mathcal P} = \prod_{n=1}^{N} {\mathcal P}_n
\end{equation}
with
\begin{eqnarray}\label{eqprojp2}
{\mathcal P}_n &=& \int_{-\infty}^{\infty} \frac{\delta \dd \alpha_n}{2 \pi}
\frac{\delta \dd \beta_{\uparrow, n}}{2 \pi} 
\frac{\delta \dd \beta_{\downarrow, n}}{2 \pi}
\int_{-\eta}^{\infty} \dd R_{e,n}  \dd R_{\uparrow, n} \dd R_{\downarrow, n} 
\nonumber\\
&\quad& \times\;  \ee^{-\ii\delta\left( (\alpha_n - \ii \lambda_0)
(R_{e,n}+R_{\uparrow, n}+R_{\downarrow, n}-1) 
-\sum_{\sigma} \beta_{\sigma, n} R_{\sigma, n}\right)}
\end{eqnarray}
which is defined on one time step $n$ only. The variable $R_{e, n}$ 
($R_{\sigma, n}$) corresponds to the amplitude of the complex $e_n$ 
($p_{\sigma, n}$) bosonic field, and $\delta \equiv \beta/N$. 
Note that it takes an infinitesimal regulator $-\eta$ for the 
integration bounds to have well defined delta functions enforcing the 
 constraints.
As previously discussed, the functional integral has to be evaluated with
$\alpha$ replaced by ($\alpha - \ii \lambda_0$) where $\lambda_0 > 0$, in
order to ensure convergence.

The fermionic contribution to the action is bi-linear, and is given for each 
spin projection by
\begin{align}\label{eq:sf}
{\mathcal S}_{\sigma}
&=  \sum_{i=1}^N \sum_{j=1}^N f^{\dagger}_{\sigma,i} [S_{\sigma}]_{i,j} f_{\sigma,j}
\nonumber
\\
&=  \sum_{n=1}^{N} 
f^{\dagger}_{\sigma,n} \Big( f_{\sigma,n} - f_{\sigma,n-1} 
\ee^{-\delta(\epsilon+\ii \beta_{\sigma, n }) } \Big) 
\end{align}
with $f_{\sigma,0} \equiv - f_{\sigma,N}$ in the second line to satisfy
anti-periodic boundary conditions. 
It yields the auxiliary fermion partition function
\begin{equation}\label{eq:zf}
 {Z}_f = Z_{\uparrow} Z_{\downarrow}
\end{equation}
with
\begin{align}\label{eq:zsigma}
{Z}_{\sigma} &= \int \!\!
 D[f^{\phantom{\dagger}}_{\sigma},f^{\dagger}_{\sigma}] \; 
 \ee^{-{\mathcal S}_{\sigma}} = \det [S_{\sigma}]
 \nonumber \\
 & =  1 + \ee^{-\beta \epsilon} \prod_{n=1}^N 
 \ee^{-\ii \delta \beta_{\sigma,n}} .  
\end{align}
The contribution to the action from the bosonic field $d$ is given by
\begin{align}\label{eq:sd}
{\mathcal S}_d
&=  \sum_{i=1}^N \sum_{j=1}^N d^{\dagger}_i [S_d]_{i,j} d_j
\nonumber
\\
&= \sum_{n=1}^{N} 
d^{\dagger}_n \Big( d_n - d_{n-1} \ee^{-\delta(U + \ii \tilde{\alpha}_n 
- \ii \beta_{\uparrow, n } - \ii \beta_{\downarrow, n }) } \Big)
\end{align}
with $d_{\sigma,0} \equiv d_{\sigma,N}$ in the second line to satisfy
periodic boundary conditions. Moreover we introduced the shorthand notation
\begin{equation}
\tilde{\alpha} \equiv \alpha - \ii \lambda_0 . 
\end{equation}
One therefore obtains
\begin{eqnarray}\label{eq:zd}
{Z}_d &=& \int \!\! D[d,d^{\dagger}] \; 
\ee^{-{\mathcal S}_d} = \det[S_d]^{-1}
\nonumber\\
&=& \left( 1 - \ee^{-\beta U} \prod_{n=1}^N \ee^{-\ii \delta 
(\tilde{\alpha}_n - \beta_{\uparrow,n} - \beta_{\downarrow,n}) } \right)^{-1} .
\end{eqnarray}
When expanding the exponential in Eqs.~(\ref{eq:sf},\ref{eq:sd}) to lowest 
order in $\delta$, the familiar form following from the Trotter-Suzuki 
decomposition is recovered. Yet, the latter may only be applied for a finite 
value of $\beta_{\sigma,n}$ while Eqs.~(\ref{eq:sf},\ref{eq:sd}) are well 
behaved even for $\beta_{\sigma,n}\rightarrow \infty$, as shown below. 

Let us now address the projection of ${Z}_f {Z}_d$ onto the physical subspace.
Two alternative procedures may be considered: i) First integrate over the radial
slave boson fields and then perform the integration over the constraint
fields, or ii) First perform the integration over the constraint fields, and
then compute the integration over the radial slave bosons. We address them in
this order.

\subsubsection{Integrating the radial bosons first}

Integrating the radial boson amplitudes yields
\begin{equation}
{\mathcal P}_n = \int_{-\infty}^{\infty}  
\frac{\dd \alpha_n}{2 \ii \pi} 
 \frac{\ee^{\ii\delta (1+\eta) \tilde{\alpha}_n}}{\tilde{\alpha}_n}
\prod_{\sigma} \frac{ \dd \gamma_{\sigma, n}}{2 \ii \pi} 
\frac{\ee^{-\ii\delta \eta \tilde{\gamma}_{\sigma,n}}}
{\tilde{\gamma}_{\sigma,n}}
\end{equation}
with
\begin{equation}
 \gamma_{\sigma,n} \equiv \beta_{\sigma,n} - \alpha_n    \:, \quad\quad 
 \tilde{\gamma}_{\sigma,n} \equiv \gamma_{\sigma,n} + \ii \lambda_0 .
\end{equation}
The Jacobian of the transformation is unity.
Then, by expanding the product ${Z}_f$, the discrete-time partition 
function is 
\begin{eqnarray}\label{eq:zn}
 {\mathcal P} {Z}_f {Z}_d &=& \int_{-\infty}^{\infty} 
 \prod_{n=1}^N \left(\frac{\dd \alpha_n}{2 \ii \pi \tilde{\alpha}_n} \prod_{\sigma} 
 \frac{ \dd \gamma_{\sigma, n}}{2 \ii \pi\tilde{\gamma}_{\sigma,n}} \right) 
 \nonumber\\
 &\;\;& \times\; \Bigg( \prod_{m=1}^N \ee^{\ii\delta (1+\eta) \tilde{\alpha}_m} 
 \ee^{-\ii\delta \eta \tilde{\gamma}_{\uparrow,m}} \ee^{-\ii\delta \eta 
 \tilde{\gamma}_{\downarrow,m}}
 \nonumber\\
 &\;\;& \;\; +\;  \ee^{-\beta \epsilon} \sum_{\sigma}  \prod_{m=1}^N 
 \ee^{\ii\delta \eta \tilde{\alpha}_m} 
 \ee^{-\ii\delta (1+\eta) \tilde{\gamma}_{\sigma,m}} 
 \ee^{-\ii\delta \eta \tilde{\gamma}_{-\sigma,m}}
 \nonumber \\
 &\;\;& \;\;+\; \ee^{-2 \beta \epsilon} \prod_{m=1}^N 
 \ee^{\ii\delta (-1+\eta) \tilde{\alpha}_m} 
 \ee^{-\ii\delta (1+\eta) \tilde{\gamma}_{\uparrow,m}} 
 \ee^{-\ii\delta (1+\eta) \tilde{\gamma}_{\downarrow,m}} \!\Bigg)
 \nonumber \\
 &\;\;& \times\; \left(1 - \ee^{-\beta U}
 \prod_{m=1}^N \ee^{\ii \delta (\tilde{\alpha}_m + \tilde{\gamma}_{\uparrow,m} 
 + \tilde{\gamma}_{\downarrow,m})} \right)^{-1} .
\end{eqnarray}

In the limit $U=\infty$ where ${Z}_d = 1$, evaluating the multiple 
integral is straightforward since it is only composed of products of simple 
integrals. The computation, however, is more intricate in the general case. 
It involves integrals of the form
\begin{eqnarray}
 I_{\xi}(r) &\equiv& \int_{-\infty}^{\infty} \frac{\dd x}{2\ii \pi (x- \ii \lambda_0)}\: 
 \frac{\ee^{\ii \delta (x - \ii \lambda_0) r}}{ 1 - \xi \ee^{\ii \delta (x - \ii \lambda_0)} }  \,, \nonumber \\
 J_{\xi}(r) &\equiv& \int_{-\infty}^{\infty} \frac{\dd x}{2\ii \pi(x + \ii \lambda_0)} \:
 \frac{\ee^{\ii \delta (x + \ii \lambda_0)r}}{ 1 - \xi \ee^{\ii \delta (x + \ii \lambda_0)} } \,,
\end{eqnarray}
where $r$ is a real number ($r \not\in -\mathbb{N}$), and $\xi$ is a complex 
exponential of other integration variables ($|\xi| <1$). Using the equalities
\begin{eqnarray}
 I_{\xi}(r) = I_0(r) + \xi I_{\xi}(r+1) &\,,& \quad I_0(r)  =  \hat{\theta}(r) \,,
 \nonumber \\
 J_{\xi}(r) = J_0(r) + \xi J_{\xi}(r+1) &\,,& \quad J_0(r)  =  -\hat{\theta}(-r) \,, 
\end{eqnarray}
where $\hat{\theta}$ is the Heaviside step function, one finds that
\begin{equation} 
 I_{\xi}(r) =  \sum_{k=0}^{\infty} \xi^k  \hat{\theta}(r+k) \;\;{\rm and}\;\;
 J_{\xi}(r) =  - \sum_{k=0}^{\infty} \xi^k \hat{\theta}(-r-k) .
\end{equation}
In particular, with $\eta \rightarrow 0^+$,
\begin{eqnarray}\label{eq:IJ}
 & I_{\xi}({-1+\eta}) = \frac{1}{1-\xi} - 1\,, \quad 
 & I_{\xi}({\eta}) = I_{\xi}({1+\eta}) = \frac{1}{1-\xi}  \,, \nonumber \\
 & J_{\xi}(-1-\eta) = -1 -\xi \,, \quad
 & J_{\xi}(-\eta) =  -1 .
\end{eqnarray}

Integrating the variables at time step $N$ is enough to separate the 
resulting multiple integral into products of simple integrals. For instance 
with the term multiplied by $\ee^{-2\beta \epsilon}$ in Eq.~(\ref{eq:zn}), 
which corresponds to the contribution from the doubly occupied state, the 
computation proceeds as follows:
\begin{eqnarray}
 &\;& \int_{-\infty}^{\infty} 
  \prod_{\sigma} \left( \frac{ \dd \gamma_{\sigma, N}}{2 \ii \pi\tilde{\gamma}_{\sigma,N}}
  \ee^{-\ii\delta (1+\eta) \tilde{\gamma}_{\sigma,N}} \right)
 \frac{\dd \alpha_N}{2 \ii \pi \tilde{\alpha}_N} 
 \ee^{\ii\delta (-1+\eta) \tilde{\alpha}_N}  {Z}_d  \nonumber \\
 &=&  \int_{-\infty}^{\infty} 
 \prod_{\sigma}\left(  \frac{ \dd \gamma_{\sigma, N}}{2 \ii \pi\tilde{\gamma}_{\sigma,N}}
 \ee^{-\ii\delta (1+\eta) \tilde{\gamma}_{\sigma,N}} \right)
  I_{\xi  \ee^{\ii \delta (\tilde{\gamma}_{\uparrow,N} + \tilde{\gamma}_{\downarrow,N})}} 
  (-1+\eta)\nonumber
\end{eqnarray}
where $\xi = \ee^{-\beta U} \prod_{m=1}^{N-1} \ee^{\ii \delta 
 (\tilde{\alpha}_m + \tilde{\gamma}_{\uparrow,m} + \tilde{\gamma}_{\downarrow,m})}$ 
\begin{eqnarray}
  &=&  \int_{-\infty}^{\infty} 
 \prod_{\sigma} \left( \frac{ \dd \gamma_{\sigma, N}}{2 \ii \pi\tilde{\gamma}_{\sigma,N}}
 \ee^{-\ii\delta (1+\eta) \tilde{\gamma}_{\sigma,N}} \right)
 \left( \frac{1}{1 - \xi  \ee^{\ii \delta (\tilde{\gamma}_{\uparrow,N} 
 + \tilde{\gamma}_{\downarrow,N})}} -1  \right)\nonumber \\
  &=&  \int_{-\infty}^{\infty} 
 \frac{ \dd \gamma_{\downarrow, N}}{2 \ii \pi\tilde{\gamma}_{\downarrow,N}}
 \ee^{-\ii\delta (1+\eta) \tilde{\gamma}_{\downarrow,N}}
 \Big( J_{\xi \ee^{\ii \delta \tilde{\gamma}_{\downarrow,N}}}(-1-\eta) 
 - J_0(-1-\eta)\Big)\nonumber \\
  &=& - \xi \int_{-\infty}^{\infty} 
 \frac{ \dd \gamma_{\downarrow, N}}{2 \ii \pi\tilde{\gamma}_{\downarrow,N}}
 \ee^{-\ii\delta \eta \tilde{\gamma}_{\downarrow,N}}  = -\xi J_0(-\eta) = \xi .
\end{eqnarray}
Hence 
\begin{eqnarray}
 {\mathcal P} {Z}_d {Z}_f &=& \int_{-\infty}^{\infty} 
 \prod_{n=1}^{N-1} \left(\frac{\dd \alpha_n}{2 \ii \pi \tilde{\alpha}_n} 
 \prod_{\sigma} \frac{ \dd \gamma_{\sigma, n}}{2 \ii \pi\tilde{\gamma}_{\sigma,n}} \right) \\
 && \times\; \Bigg( \prod_{m=1}^{N-1} \ee^{\ii\delta (1+\eta) \tilde{\alpha}_m} 
 \ee^{-\ii\delta \eta \tilde{\gamma}_{\uparrow,m}} \ee^{-\ii\delta \eta \tilde{\gamma}_{\downarrow,m}}
 \nonumber\\
 && \quad + \; \ee^{-\beta \epsilon} \sum_{\sigma}  \prod_{m=1}^{N-1} \ee^{\ii\delta \eta \tilde{\alpha}_m} 
 \ee^{-\ii\delta (1+\eta) \tilde{\gamma}_{\sigma,m}} \ee^{-\ii\delta \eta \tilde{\gamma}_{-\sigma,m}}
 \nonumber \\
 && \quad + \; \ee^{-2 \beta \epsilon} \ee^{-\beta U} \prod_{m=1}^{N-1} \ee^{\ii\delta \eta \tilde{\alpha}_m} 
 \ee^{-\ii\delta \eta \tilde{\gamma}_{\uparrow,m}} 
 \ee^{-\ii\delta \eta \tilde{\gamma}_{\downarrow,m}} \Bigg) \nonumber .
\end{eqnarray}
Then integrating over the remaining $(N-1)$ time steps is straightforward, 
and
\begin{eqnarray}
{\mathcal P} {Z}_d {Z}_f &=&  \left( I_0({1+\eta}) J_0({-\eta})^2 \right)^{N-1}
 \nonumber \\
 &\quad& +\; 2 \ee^{-\beta \epsilon} \left( I_0({\eta}) J_0({-1-\eta}) J_0({-\eta}) \right)^{N-1}
 \nonumber \\
 &\quad& +\; \ee^{-\beta (2 \epsilon + U)} \left( I_0({\eta}) J_0({-\eta})^2 \right)^{N-1}
 \nonumber \\
 &=&  1 + 2 \ee^{-\beta \epsilon}  + \ee^{-\beta (2 \epsilon + U)}.
\end{eqnarray}

Even though this procedure might be seen as lacking transparency, the proper 
result is obtained. Let us now consider an alternative route along which one 
first performs the integrals over the Lagrange multipliers.

\subsubsection{Integrating the Lagrange multipliers first}
Integrating first the Lagrange multipliers involves evaluating integrals of 
the form
\begin{equation}
 \int_{-\infty}^{\infty} \frac{\delta \dd x}{2\pi}\: 
 \frac{\ee^{\ii \delta (x - \ii \lambda_0) r}}{ 1 - \xi \ee^{\ii \delta (x - \ii \lambda_0)} } 
  = \frac{\partial I_{\xi}(r) }{\partial r}  
  =  \sum_{k=0}^{\infty} \xi^k \hat{\delta}(r+k) 
\end{equation}
where $\hat{\delta}$ is the Dirac delta function. As shown below, the relation 
allows to write the partition function as an explicit sum over the different 
values of the double occupancy $|d|^2$. Hence, expanding ${Z}_d$ as 
a power series of $\ee^{-\beta U}$ yields
\begin{eqnarray}
&&{\mathcal P} {Z}_f {Z}_d 
\nonumber\\
&&\; = {\mathcal P} \prod_{\sigma}\!
\left(\! 1 + \ee^{-\beta \epsilon} \ee^{-\ii \delta \!\sum\limits_m \beta_{\sigma,m}} \right)
 \!\!\sum_{k=0}^{\infty}\! \left(\! \ee^{-\beta U} \ee^{\ii \delta\! \sum\limits_m (-\tilde{\alpha}_m
+ \beta_{\uparrow,m} + \beta_{\downarrow,m})} \right)^{\!\! k}
\nonumber\\
&&\; = \prod_{p=1}^N \bigg( \int_{-\eta}^{\infty} \!
\dd R_{e,p}  \dd R_{\uparrow, p} \dd R_{\downarrow, p} 
\int_{-\infty}^{\infty} \frac{\delta \dd \alpha_p}{2 \pi}
\frac{\delta \dd \beta_{\uparrow, p}}{2 \pi} \frac{\delta \dd \beta_{\downarrow, p}}{2 \pi} \bigg)
\nonumber\\
&&\quad \;\times\; \Bigg\{  \sum_{k=0}^{\infty} \ee^{-k\beta  U}  
\prod_{n=1}^N \ee^{\ii \delta(1-k-R_{e,n}-R_{\uparrow, n}-R_{\downarrow, n}) \tilde{\alpha}_n} 
\nonumber\\
&&\quad \quad \times \; \bigg(  \prod_{m=1}^N \ee^{\ii \delta (R_{\uparrow,m}+k)\beta_{\uparrow,m}}
\ee^{\ii \delta (R_{\downarrow,m}+k)\beta_{\downarrow,m}} 
\nonumber\\
&&\quad\quad\quad + \; \ee^{-\beta\epsilon} \sum_{\sigma} \prod_{m=1}^N \ee^{\ii \delta (R_{\sigma,m}+k)\beta_{\sigma,m}}
\ee^{\ii \delta (R_{-\sigma,m}+k-1)\beta_{-\sigma,m}} 
\nonumber\\
&&\quad\quad\quad + \; \ee^{-2\beta\epsilon} \prod_{m=1}^N \ee^{\ii \delta (R_{\uparrow,m}+k-1)\beta_{\uparrow,m}}
\ee^{\ii \delta (R_{\downarrow,m}+k-1)\beta_{\downarrow,m}} \!\bigg) \Bigg\} .
\end{eqnarray}
Here one recognizes the Fourier transform of delta functions. So integrating
the Lagrange multipliers results in
\begin{eqnarray}
{\mathcal P} {Z}_f {Z}_d &=& \prod_{p=1}^N \bigg( \int_{-\eta}^{\infty}
\dd R_{e,p}  \dd R_{\uparrow, p} \dd R_{\downarrow, p} \bigg) 
\\
&& \times \Bigg\{\! \sum_{k=0}^{\infty} \ee^{-k\beta  U} 
\prod_{n=1}^N \hat{\delta }\left( 1-k-R_{e,n}\!-R_{\uparrow, n}\!-R_{\downarrow, n}\right)
\nonumber\\
&& \;\; \times \bigg( \prod_{m=1}^N \hat{\delta}(R_{\uparrow,m}\!+k) \hat{\delta}(R_{\downarrow,m}\!+k) 
\nonumber\\
&& \quad + \; \ee^{-\beta\epsilon} \sum_{\sigma} \prod_{m=1}^N \hat{\delta}(R_{\sigma,m}\!+k)
\hat{\delta}(R_{-\sigma,m}\!+k-1) 
\nonumber\\
&& \quad + \; \ee^{-2\beta\epsilon} \prod_{m=1}^N \hat{\delta}(R_{\uparrow,m}\!+k-1)
\hat{\delta}(R_{\downarrow,m}\!+k-1) \! \bigg) \!\Bigg\} . \nonumber
\end{eqnarray}
As stated above, the partition function is obtained as a sum running over 
the integer values $k$ of the boson amplitude $|d|^2$. Then integrating the 
delta function $\hat{\delta}(R_{\sigma,m}+k)$ yields a non vanishing 
contribution only for $k=0$ because $R_{\sigma,m} \geq -\eta$. With 
$\hat{\delta}(R_{\sigma,m}+k-1)$, only $k=0$ and $k=1$ are retained. Thus 
the sum stops at $k=1$, and
\begin{eqnarray}
{\mathcal P} {Z}_f {Z}_d &=& \prod_{p=1}^N 
\bigg( \int_{-\eta}^{\infty}\!\! \dd R_{e,p} \bigg) 
\Bigg\{\!  \bigg(\! \prod_{n=1}^N \hat{\delta }\left( 1-R_{e,n}\right) 
\nonumber\\
&\quad& + \; 2 \ee^{-\beta\epsilon} \prod_{n=1}^N \hat{\delta }\left(-R_{e,n}\right) 
+ \ee^{-2\beta\epsilon} \prod_{n=1}^N \hat{\delta }\left(-1-R_{e,n}\right) \!\!\bigg)
\nonumber\\
&\quad& + \; \ee^{-\beta  U} \bigg( 0  + \ee^{-\beta\epsilon}\times 0 
+ \ee^{-2\beta\epsilon} \prod_{n=1}^N \hat{\delta }\left(-R_{e,n}\right) 
\! \bigg) \! \Bigg\}
\nonumber\\
&=& 1 + 2\ee^{-\beta \epsilon}  + \ee^{-\beta (2 \epsilon + U)}.
\end{eqnarray}

The second procedure is eventually more intuitive than the first one, as the
projection results in an explicit sum over the contribution of every physical
state at every time step. That is why we will use it to discuss the evaluation
of the Green's function. 

\section{Green's function}
For the imaginary-time variable $0<\tau<\beta$, the time-ordered Green's
function is calculated as the limit of the discrete-time correlation:
\begin{equation}
  {\mathcal Z} {\mathcal G}_{\uparrow}(\tau,0)  
  = -\langle c_{\uparrow}^{\phantom{\dagger}}(\tau) c_{\uparrow}^{\dagger}(0) \rangle
  = -\lim_{N\rightarrow \infty}
  \langle c_{\uparrow,m}^{\phantom{\dagger}}  c_{\uparrow,1}^{\dagger} \rangle 
\end{equation}
with $\lim\limits_{N\rightarrow \infty}\left(\frac{m \beta}{N}\right)
=\tau$. Using the KR representation, Eq.~(\ref{eq:KR}), one obtains the
correlation $\langle c_{\uparrow,m}^{\phantom{\dagger}}  c_{\uparrow,1}^{\dagger} \rangle $
as the sum of the formal projections $P$ onto the physical subspace of the
correlations of two kinds of processes within the augmented Fock space of
auxiliary boson states: 
 \begin{equation}\label{eq:corr}
  P\left[ \big(e^{\dagger} p_{\uparrow}^{\phantom{\dagger}} f_{\uparrow}\big)_m
  \big(p_{\uparrow}^{\dagger} e f_{\uparrow}^{\dagger}\big)_1 \right]
  + P\left[ \big(p_{\downarrow}^{\dagger} d f_{\uparrow}^{\phantom{\dagger}}\big)_m   
  \big(d^{\dagger} p_{\downarrow}^{\phantom{\dagger}}  f_{\uparrow}^{\dagger}\big)_1 
  \right] .
\end{equation}
The first process involves variations of 
$e$ and $p_{\uparrow}$ boson numbers. When restricted to the physical subspace, it 
generates the transitions $|0\rangle \rightarrow |\!\uparrow \rangle \rightarrow |0\rangle$
between the physical states, while the second process, with variations of $d$ and 
$p_{\downarrow}$ boson numbers, corresponds to the transitions 
$|\!\downarrow \rangle \rightarrow |\!\uparrow\downarrow \rangle \rightarrow 
|\!\downarrow\rangle$. 

\begin{figure}
  \includegraphics[clip=true, width=0.47\textwidth]
  {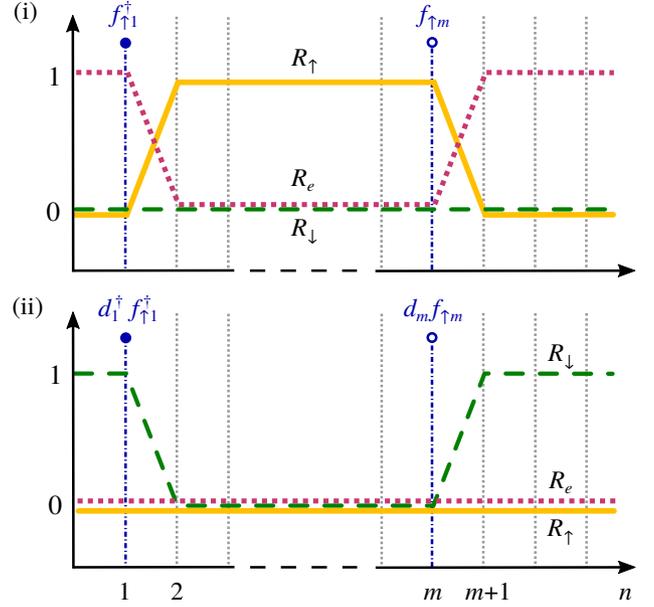}
  \caption{(Color online) Time evolution of constrained radial fields during the 
  transitions (i) $|0\rangle \rightarrow |\!\uparrow \rangle \rightarrow |0\rangle$, and 
  (ii) $|\!\downarrow \rangle \rightarrow |\!\uparrow\downarrow \rangle \rightarrow 
  |\!\downarrow\rangle$, generated by the creation of an electron with spin up 
  at time step~1, and its annihilation at time step~$m>1$. The values of $R_{\uparrow}$, 
  $R_{\downarrow}$, and $R_e$ are plotted respectively with solid, dashed, and 
  dotted lines.}
  \label{fig:evo-1}
\end{figure}

Calculating functional integrals with the Cartesian representation of complex bosonic 
fields is standard~\cite{negele}. For instance, at time $\tau_n=\frac{n\beta}{N}$, the operators 
$e$ and $e^{\dagger}$ yield, respectively, the complex fields $e_{n-1}$ and $e_n^{\dagger}$ in
the discrete-time action. So the number operator $e^{\dagger}e$ results in the 
value $e_n^{\dagger} e_{n-1}$. 
For the radial representation, an original procedure has to be devised~\cite{RFTK01}.
It has been found that the latter contribution to the action can be computed with the 
real field $R_{e,n}$, which describes the boson amplitude $|e|^2$. However, in the 
correlation Eq.~(\ref{eq:corr}) the annihilation and creation operators act at different 
times. The issue is then how to integrate terms involving a single operator $e$ or 
$e^{\dagger}$. One could imagine having the license to use either $\sqrt{R_{e,n-1}}$ or 
$\sqrt{R_{e,n}}$. However, this is not the case. As shown in details below, when the 
dynamics of the $d$-boson and $f$-fermion fields are imposed, the physical 
constraints~(\ref{eqrf:q2_KR}) strictly select the radial-field trajectories that contribute 
to the functional integral. 
These constrained evolutions are plotted in Fig.~\ref{fig:evo-1} for the two processes 
generated by the creation of an electron at time step~1, followed by its annihilation at 
time step~$m$.
Hence, in order to obtain non-vanishing correlations and the correct result for the 
Green's function, the proper choice of time steps for the radial fields is necessarily
\begin{align}
 \big(e^{\dagger} p_{\uparrow}^{\phantom{\dagger}} f_{\uparrow}\big)_m
  \big(p_{\uparrow}^{\dagger} e f_{\uparrow}^{\dagger}\big)_1 
  &\leadsto
 \sqrt{R_{e,m+1}R_{\uparrow,m}} f_{\uparrow,m}
 \sqrt{R_{\uparrow,2}^{\phantom{\dagger}} R_{e,1}} f_{\uparrow,1}^{\dagger}
 \nonumber \\
 \big(p_{\downarrow}^{\dagger} d f_{\uparrow}^{\phantom{\dagger}}\big)_m   
  \big(d^{\dagger} p_{\downarrow}^{\phantom{\dagger}}  f_{\uparrow}^{\dagger}\big)_1 
 &\leadsto
 \sqrt{R_{\downarrow,m+1}} d_{m} f_{\uparrow,m} d_1^{\dagger} 
 \sqrt{R_{\downarrow,1}}  f_{\uparrow,1}^{\dagger} .
 \end{align}
This is obtained with the representations for the electron annihilation and 
creation fields
\begin{equation}\label{eq:c_rad_KR}
  \begin{array}{l}
          c_{\uparrow,n} \leadsto \left( \sqrt{R_{e,n+1} R_{\uparrow,n}}  
          + \sqrt{R_{\downarrow,n+1}} d_n \right) f_{\uparrow,n}^{\phantom{\dagger}}
          \\
          c_{\uparrow,n}^{\dagger} \leadsto \left( \sqrt{R_{\uparrow,n+1} R_{e,n}}  
          + d_n^{\dagger} \sqrt{R_{\downarrow,n}}  \right) f_{\uparrow,n}^{\dagger}
         \end{array}    
\end{equation}
which are the generalization for finite on-site repulsion of
the forms obtained in~\cite{RFTK01} for the Barnes representation, and in
\cite{RFTK12} for the KR representation in the limit of
infinite~$U$. Eq.~(\ref{eq:c_rad_KR}) also serves to express the hybridization
term of the single-impurity Anderson Model within this auxiliary particles
framework. 
 
Specifically, the projected correlation for the first process 
may be computed using Eqs.~(\ref{eqprojp},\ref{eqprojp2}) as
\begin{align}
 & P\Big[\big(e^{\dagger} p_{\uparrow}^{\phantom{\dagger}} f_{\uparrow}\big)_m
  \big(p_{\uparrow}^{\dagger} e f_{\uparrow}^{\dagger}\big)_1 \Big]
 \nonumber \\ 
 & = 
 \lim_{\eta\rightarrow 0^{+}} {\mathcal P} 
 \left\langle \sqrt{R_{e,m+1}R_{\uparrow,m}} f_{\uparrow,m}
 \sqrt{R_{\uparrow,2}^{\phantom{\dagger}} R_{e,1}} f_{\uparrow,1}^{\dagger} \right\rangle
 \nonumber \\
 & = 
 \lim_{\eta\rightarrow 0^{+}} {\mathcal P} 
 \left( \sqrt{R_{e,m+1}R_{\uparrow,m}R_{\uparrow,2} R_{e,1}} Z_d 
 \left\langle f_{\uparrow,m}^{\phantom{\dagger}} f_{\uparrow,1}^{\dagger} \right\rangle Z_{\downarrow} \right) . 
\end{align}
The angle brackets in the second line stand for the integration over the trajectories of 
the $f_{\sigma}$ and $d$ fields, which yields the partition functions given by 
Eqs.~(\ref{eq:zsigma},\ref{eq:zd}), and  
\begin{align}
 \left\langle f_{\uparrow,m}^{\phantom{\dagger}} f_{\uparrow,1}^{\dagger} \right\rangle
  &=\int \!\!\! D[f^{\phantom{\dagger}}_{\uparrow},f^{\dagger}_{\uparrow}] \; 
 f_{\uparrow,m}^{\phantom{\dagger}} f_{\uparrow,1}^{\dagger} \; \ee^{-{\mathcal S}_{\uparrow}}
 \nonumber \\
  &= [S_{\uparrow}^{-1}]_{m,1} \det [S_{\uparrow}] 
  = \prod_{n=2}^{m} \ee^{-\delta(\epsilon + \ii \beta_{\uparrow,n})} .
\end{align}  
As in the previous section, expanding $Z_d$ in power series of $\ee^{-\beta U}$
results in a sum over the integer values $k$ of $|d|^2$. The latter actually stops
 at $k=0$. Indeed, integrating out the Lagrange multipliers Fourier transforms the 
 complex exponentials into delta functions, and it turns out that every term in the 
 sum contains a product $\prod\limits_{n=2}^m \hat{\delta}(R_{\uparrow,n} + k)$. Since 
 $R_{\uparrow,n} \ge 0$, only $k=0$ actually produces a non vanishing contribution. 
 Thus the projected correlation can be written as
\begin{align}
 & P\Big[\big(e^{\dagger} p_{\uparrow}^{\phantom{\dagger}} f_{\uparrow}\big)_m
  \big(p_{\uparrow}^{\dagger} e f_{\uparrow}^{\dagger}\big)_1 \Big]
   = \ee^{-\delta(m-1) \epsilon}   \prod_{n=1}^N \left( \int_{0^-}^{+\infty} 
   \!\!\!\!\! dR_{e,n}\, \prod_{\sigma} dR_{\sigma,n} \right)
 \nonumber \\
 &\times \sqrt{R_{e,m+1}R_{\uparrow,m}R_{\uparrow,2} R_{e,1}} \; 
 \prod_{n_1=1}^N \hat{\delta} \big( R_{e,n_1} + R_{\uparrow,n_1} + R_{\downarrow,n_1} - 1 \big)
 \nonumber \\
 & \times \left( \prod_{n_2=1}^N \hat{\delta}(R_{\downarrow,n_2}) 
 + \ee^{-\beta \epsilon} \prod_{n_2=1}^N \hat{\delta}(R_{\downarrow,n_2}-1) \right)
 \nonumber \\
 & \times \prod_{i_1 \in \llbracket 2,m \rrbracket} \hat{\delta}(R_{\uparrow,i_1}-1) 
 \prod_{i_2 \notin \llbracket 2,m \rrbracket} \hat{\delta}(R_{\uparrow,i_2})
\end{align}
(in the product, ${i_2 \notin \llbracket 2,m \rrbracket}$ is a shorthand notation for 
the integer values $i_2=1$ or $m+1 \le i_2 \le N$).
The expression shows that only one trajectory of the radial fields contribute to the
functional integral. This is the constrained evolution depicted in 
Fig.~\ref{fig:evo-1}, in which $R_{\uparrow,n}= 1$ for $2\le n\le m$ and $R_{\uparrow,n}=0$ 
otherwise, $R_{e,n} = 1 - R_{\uparrow,n}$ and $R_{\downarrow,n}=0$. And, as a result,
\begin{equation}
 P\Big[\big(e^{\dagger} p_{\uparrow}^{\phantom{\dagger}} f_{\uparrow}\big)_m
  \big(p_{\uparrow}^{\dagger} e f_{\uparrow}^{\dagger}\big)_1 \Big]
   = \ee^{-\delta(m-1) \epsilon}.
\end{equation}

For the second process,
\begin{align}
 & P\Big[\big(p_{\downarrow}^{\dagger} d f_{\uparrow}^{\phantom{\dagger}}\big)_m   
  \big(d^{\dagger} p_{\downarrow}^{\phantom{\dagger}}  f_{\uparrow}^{\dagger}\big)_1 \Big]
 \nonumber \\ 
 &= \lim_{\eta\rightarrow 0^{+}} {\mathcal P} 
 \left\langle \sqrt{R_{\downarrow,m+1}} d_{m} f_{\uparrow,m} d_1^{\dagger} 
 \sqrt{R_{\downarrow,1}}  f_{\uparrow,1}^{\dagger} \right\rangle
 \nonumber \\
 & = \lim_{\eta\rightarrow 0^{+}} {\mathcal P} 
 \left( \sqrt{R_{\downarrow,m+1}R_{\downarrow,1}} 
 \left\langle d_{m}^{\phantom{\dagger}} d_1^{\dagger} \right\rangle 
  \left\langle f_{\uparrow,m}^{\phantom{\dagger}} f_{\uparrow,1}^{\dagger} \right\rangle 
  Z_{\downarrow} \right)
\end{align}
where $ \left\langle d_{m}^{\phantom{\dagger}} d_1^{\dagger} \right\rangle $
is obtained as 
\begin{align}
 \left\langle d_{m}^{\phantom{\dagger}} d_1^{\dagger} \right\rangle
 &= \int \!\!\! D[d,d^{\dagger}] \;\; 
 d_{m}^{\phantom{\dagger}} d_1^{\dagger} \; \ee^{-{\mathcal S}_d}
 \nonumber \\
 &= \frac{[S_d^{-1}]_{m,1}}{\det [S_d]} 
 = Z_d^2 \prod\limits_{n=2}^{m} \ee^{-\delta\big(U+ \ii (\tilde{\alpha}_n - \beta_{\uparrow,n}
 - \beta_{\downarrow,n})\big)} .
\end{align}
Expanding now $Z_d^2$ and integrating out the Lagrange multipliers result in
\begin{align}
 & P\Big[\big(p_{\downarrow}^{\dagger} d f_{\uparrow}^{\phantom{\dagger}}\big)_m   
  \big(d^{\dagger} p_{\downarrow}^{\phantom{\dagger}}  f_{\uparrow}^{\dagger}\big)_1 \Big]
 \nonumber \\  
  &  = \ee^{-\delta(m-1)(U+\epsilon)} 
  \prod_{n=1}^N \left( \int_{0^-}^{+\infty} 
   \!\!\!\!\! dR_{e,n}\, \prod_{\sigma} dR_{\sigma,n} \right) 
   \sqrt{R_{\uparrow,m+1}R_{\downarrow,1}} 
 \nonumber \\
 &\times 
 \prod_{i_1=2}^m \!\hat{\delta} \big( R_{e,i_1} + R_{\uparrow,i_1} 
 + R_{\downarrow,i_1} \big)
 \prod_{i_2\notin \llbracket 2,m \rrbracket} \!\!\!\!\!\hat{\delta} 
 \big( R_{e,i_2} + R_{\uparrow,i_2} + R_{\downarrow,i_2} - 1 \big)
 \nonumber \\
 &\times \prod_{n_1=1}^N \hat{\delta}(R_{\uparrow,n_1}) 
  \times \left( \prod_{j_1=2}^m \hat{\delta}(R_{\downarrow,j_1}+1)
  \prod_{j_2\notin \llbracket 2,m \rrbracket} \hat{\delta}(R_{\downarrow,j_2}) \right.
 \nonumber \\ 
 &\quad \left. + \ee^{-\beta \epsilon} 
 \prod_{l_1=2}^m \hat{\delta}(R_{\downarrow,l_1}) 
 \prod_{l_2\notin \llbracket 2,m \rrbracket} \hat{\delta}(R_{\downarrow,l_2}-1)\right).
\end{align}
Among all the trajectories, only the second one shown in Fig.~\ref{fig:evo-1} yields
a non-vanishing contribution: the delta functions impose that $R_{\downarrow,n}= 0$ for 
$2\le n\le m$ and $R_{\downarrow,n}=1$ otherwise, while $R_{\uparrow,n} =
R_{e,n} =0$ for each time step. Hence after integrating over the radial
variables, the second projected correlation 
is
\begin{equation}
 P\Big[\big(p_{\downarrow}^{\dagger} d f_{\uparrow}^{\phantom{\dagger}}\big)_m   
  \big(d^{\dagger} p_{\downarrow}^{\phantom{\dagger}}  f_{\uparrow}^{\dagger}\big)_1 \Big]
 = \ee^{-\delta(m-1)(U+\epsilon)}  \ee^{-\beta \epsilon} .   
\end{equation}

Lastly, in the limit $N\rightarrow\infty$, the expected expression for the Green's 
function is obtained as
\begin{equation}
  {\mathcal Z} {\mathcal G}_{\uparrow}(\tau,0) 
  = -\left( \ee^{-\tau \epsilon} + \ee^{-\tau(U+\epsilon)} \ee^{-\beta
      \epsilon} \right).
\end{equation}

\begin{figure}
  \includegraphics[clip=true, width=0.47\textwidth]
  {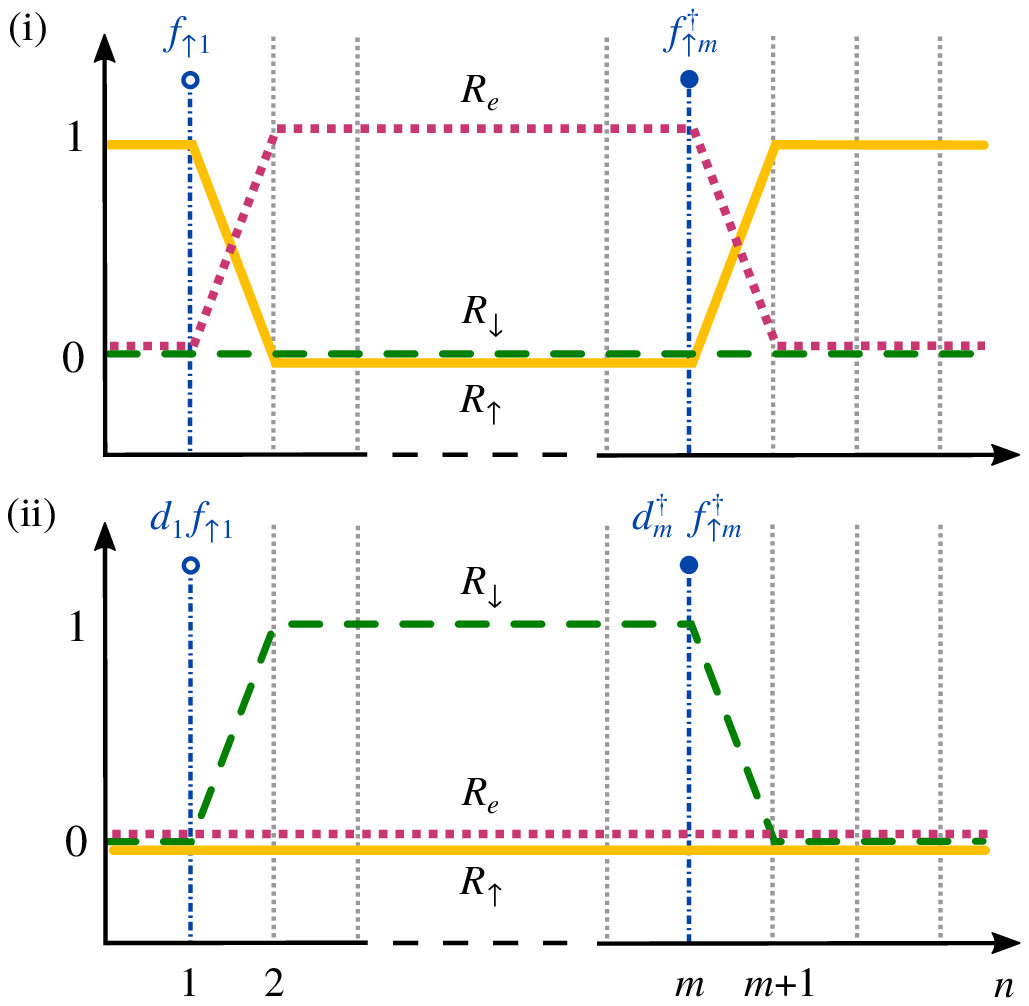}
  \caption{(Color online) Time evolution of constrained radial fields during the 
  transitions (i) 
  $|\!\uparrow \rangle \rightarrow  |0\rangle \rightarrow |\!\uparrow \rangle$, and 
  (ii) $ |\!\uparrow\downarrow \rangle \rightarrow |\!\downarrow \rangle \rightarrow 
  |\!\uparrow\downarrow \rangle$, generated by the annihilation of an electron with
  spin up at time step~1, and its creation at time step~$m>1$. The values of $R_{\uparrow}$, 
  $R_{\downarrow}$, and $R_e$ are plotted respectively with solid, dashed, and dotted 
  lines.}
  \label{fig:evo-2}
\end{figure}

The same procedure can be applied to calculate the time-ordered Green's function
\begin{equation}
  {\mathcal Z} {\mathcal G}_{\uparrow}(0,\tau)  
  = \langle c_{\uparrow}^{\dagger}(\tau)  c_{\uparrow}^{\phantom{\dagger}}(0) \rangle
  = \lim_{N\rightarrow \infty}
  \langle   c_{\uparrow,m}^{\dagger}  c_{\uparrow,1}^{\phantom{\dagger}}\rangle .
\end{equation}
Here the annihilation of an electron with spin up at time step 1, followed by the 
creation of another one at $m>1$, produces two possible series of transitions between
the physical states: 
$|\!\uparrow \rangle \rightarrow |0\rangle \rightarrow |\!\uparrow \rangle$, and 
$|\!\uparrow \downarrow \rangle \rightarrow |\!\downarrow \rangle
\rightarrow|\!\uparrow \downarrow \rangle$. The correlation 
$\langle   c_{\uparrow,m}^{\dagger}  c_{\uparrow,1}^{\phantom{\dagger}}\rangle$ is then 
the sum of the projected correlations
\begin{align}
 & P\Big[\big(p_{\uparrow}^{\dagger} e f_{\uparrow}^{\dagger} \big)_m \big(e^{\dagger} 
 p_{\uparrow}^{\phantom{\dagger}} f_{\uparrow}^{\phantom{\dagger}} \big)_1 \Big]
 \nonumber \\ 
 & = 
 \lim_{\eta\rightarrow 0^{+}} {\mathcal P} 
 \left\langle \sqrt{ R_{\uparrow,m+1}^{\phantom{\dagger}}  R_{e,m}^{\phantom{\dagger}} }  
 f_{\uparrow,m}^{\dagger} 
 \sqrt{ R_{e,2}^{\phantom{\dagger}} R_{\uparrow,1}^{\phantom{\dagger}} } 
 f_{\uparrow,1}^{\phantom{\dagger}}
 \right\rangle
\end{align}
and
\begin{align}
 & P\Big[\big(d^{\dagger} p_{\downarrow}^{\phantom{\dagger}} f_{\uparrow}^{\dagger} \big)_m
 \big(p_{\downarrow}^{\dagger} d^{\phantom{\dagger}} f_{\uparrow}^{\phantom{\dagger}}\big)_1 \Big]
 \nonumber \\ 
 &= \lim_{\eta\rightarrow 0^{+}} {\mathcal P} 
 \left\langle  \sqrt{R_{\downarrow,m}^{\phantom{\dagger}}} d_{m}^{\phantom{\dagger}} 
 f_{\uparrow,m}^{\dagger} 
 \sqrt{R_{\downarrow,2}^{\phantom{\dagger}}} d_1^{\phantom{\dagger}} 
 f_{\uparrow,1}^{\phantom{\dagger}} \right\rangle,
\end{align}
with
\begin{equation}
 \left\langle f_{\uparrow,m}^{\dagger} f_{\uparrow,1}^{\phantom{\dagger}} \right\rangle
  = \prod_{n\notin \llbracket 2,m \rrbracket} \ee^{-\delta(\epsilon + \ii \beta_{\uparrow,n})}
\end{equation} 
and
\begin{equation}
 \left\langle d_{m}^{\dagger} d_1^{\phantom{\dagger}} \right\rangle
 = Z_d^2 \prod_{n\notin \llbracket 2,m \rrbracket}  \ee^{-\delta\big(U+ \ii (\tilde{\alpha}_n - \beta_{\uparrow,n}
 - \beta_{\downarrow,n})\big)}.
\end{equation}
This radial representation of the correlations is obtained from the
discrete-time prescriptions~(\ref{eq:c_rad_KR}) as well. It is the only choice
of time-steps that yields non-vanishing integrals, as illustrated by
Fig.~\ref{fig:evo-2} which displays the constrained evolutions of the radial
fields for the physical transitions involved here. Then
projecting the sum onto the physical subspace, that is integrating out the
Lagrange multipliers and the radial boson fields, results in the expected
correlation for the Green's function 
\begin{align}
  {\mathcal Z} {\mathcal G}_{\uparrow}(0,\tau)  
  & = \lim_{N\rightarrow \infty}
 \left(  \ee^{-\delta(N-m+1)\epsilon} + \ee^{-\beta \epsilon} 
 \ee^{-\delta(N-m+1)(\epsilon+U)}\right)
  \nonumber \\
  & = \ee^{-(\beta - \tau)\epsilon} + \ee^{-\beta \epsilon} 
 \ee^{-(\beta -\tau)(\epsilon+U)} .
\end{align}
Higher order Green's functions may be calculated by means of the same procedure.

\section{Summary and conclusion}\label{sec:conclusion}

In this work, we have considered the Kotliar and Ruckenstein slave boson
representation of the Hubbard Model in the radial gauge. It allows to gauge
away the phase of three of the four involved slave boson fields. As a result,
the functional integral representation of the partition function, of the Green's
function and of correlation functions involves canonical fermionic and bosonic
fields, together with radial slave boson fields. The correctness of this
functional integral representation has been verified through the exact
calculation of the partition function and the Green's function in the atomic
limit. Furthermore, the proper representation of the physical electron
creation and annihilation operators, that is crucial to the representation of
the kinetic energy operator, has been established. This paves the way for
calculations of larger systems and can also be applied to 
the single-impurity Anderson model.


\begin{thebibliography}{99}


\bibitem{Barnes76}S.\,E. Barnes, 
\textit{J. Phys. F} \textbf{1976}, \textit{6}, 1375.

\bibitem{Barnes77}S.\,E. Barnes, 
\textit{J. Phys. F} \textbf{1977}, \textit{7}, 2637.

\bibitem{KR}G.~Kotliar and A.\,E. Ruckenstein, 
\textit{Phys. Rev. Lett.} \textbf{1986}, \textit{57}, 1362.

\bibitem{FK} R.~Fr\'esard and G.~Kotliar, 
\textit{Phys. Rev. B} \textbf{1997}, \textit{56}, 12909.

\bibitem{LiWH}T.\,C. Li, P.~W\"olfle, and P.\,J. Hirschfeld, 
             \textit{Phys. Rev. B} \textbf{1989}, \textit{40}, 6817.

\bibitem{FW92} R. Fr\'esard and P. W\"olfle,
              \textit{Int. J. of Mod. Phys. B} \textbf{1992}, \textit{6},  685;
                                         \textbf{1992}, \textit{6}, 3087.

\bibitem{Kotliar-Georges} F.~Lechermann, A.~Georges, G.~Kotliar, and
O.~Parcollet, \textit{Phys. Rev. B} \textbf{2007}, \textit{76}, 155102.

\bibitem{Piefke} C.~Piefke and F.~Lechermann,
            \textit{Phys. Rev. B} \textbf{2018}, \textit{97}, 125154.

\bibitem{Ste17} K. Steffen, R. Fr\'esard, and T. Kopp, 
                   \textit{Phys. Rev. B} \textbf{2017}, \textit{95}, 035143. 

\bibitem{RN83a} N.~Read, D.\,M. Newns, 
               \textit{J. Phys. C} \textbf{1983}, \textit{16}, L1055.

\bibitem{RN83b} N.~Read, D.\,M. Newns, 
               \textit{J. Phys. C} \textbf{1983}, \textit{16}, 3273.
  
\bibitem{NR87} D.\,M. Newns and N.~Read, 
               \textit{Adv. in Physics} \textbf{1987}, \textit{36}, 799.

\bibitem{RFTK01} R.~Fr\'esard and T.~Kopp, 
                \textit{Nucl. Phys. B} \textbf{2001}, \textit{594}, 769.

\bibitem{RFHOTK07} R.~Fr\'esard, H. Ouerdane, and T. Kopp, 
                   \textit{Nucl. Phys. B} \textbf{2007}, \textit{785}, 286.

\bibitem{Lil90} L. Lilly, A. Muramatsu, and W. Hanke,
                    \textit{Phys. Rev. Lett.} \textbf{1990}, \textit{65}, 1379.

\bibitem{Fre91} R. Fr\'esard, M. Dzierzawa, and P. W\"olfle,
                    \textit{Europhys. Lett.} \textbf{1991}, \textit{15}, 325.

\bibitem{Igo13} P.~A. Igoshev, M.~A. Timirgazin, A.~K. Arzhnikov, and
  V.~Y. Irkhin,  \textit{JETP Lett.} \textbf{2013}, \textit{98}, 150.
  
\bibitem{Fre92} R. Fr\'esard and P. W\"olfle,
                \textit{J. Phys.: Condens. Matter} \textbf{1992}, \textit{4}, 3625.
                    
\bibitem{Doll2} B. M\"oller, K. Doll, and R. Fr\'esard,
                \textit{J. Phys.: Condens. Matter} \textbf{1993}, \textit{5}, 4847.

\bibitem{SeiSi} G. Seibold, E. Sigmund, and V. Hizhnyakov,
                \textit{Phys. Rev. B} \textbf{1998}, \textit{57}, 6937.    

\bibitem{Fle01} M. Fleck, A. I. Lichtenstein, and A.~M. Ole\'s,
                \textit{Phys. Rev. B} \textbf{2001}, \textit{64}, 134528.


\bibitem{Sei02} J. Lorenzana and G. Seibold,
                   \textit{Phys. Rev. Lett.} \textbf{2002}, \textit{89}, 136401;
                                     \textbf{2003}, \textit{90}, 066404;
                                      \textbf{2005}, \textit{94}, 107006.

\bibitem{Rac06a} M. Raczkowski, R.~Fr\'esard, and A.~M. Ole\'s,
                    \textit{Phys. Rev. B} \textbf{2006}, \textit{73}, 174525.

\bibitem{Rac06b}    M. Raczkowski, M. Capello, D. Poilblanc, R. Fr\'esard,
                    and A. M. Ole\'s,
                    \textit{Phys. Rev. B} \textbf{2007}, \textit{76}, 140505(R).

\bibitem{RaEPL} M. Raczkowski, R.~Fr\'esard, and A.~M.~Ole\'s,
                    \textit{Europhys. Lett.} \textbf{2006}, \textit{76}, 128.

\bibitem{Fre02} R. Fr\'esard and M. Lamboley, 
                  \textit{J. Low Temp. Phys.} \textbf{2002}, \textit{126}, 1091.
                    
\bibitem{lhoutellier15} G. Lhoutellier, R. Fr\'esard, and A. M. Ole\'s, 
                       \textit{Phys. Rev. B} \textbf{2015}, \textit{91}, 224410.
                        
\bibitem{FW98} R. Fr\'esard and W. Zimmermann, 
                   \textit{Phys. Rev. B} \textbf{1998}, \textit{58}, 15288.

\bibitem{Igo15} P.A. Igoshev, M.A. Timirgazin, V.F. Gilmutdinov,
                A.K. Arzhnikov, V.Yu. Irkhin,
           \textit{J. Phys: Condens. Matter} \textbf{2015}, \textit{27}, 446002.

\bibitem{Ras88} J.~W. Rasul and T. Li,
       \textit{J. Phys. C: Solid State Phys.} \textbf{1988}, \textit{21}, 5119.

                    
\bibitem{Lav90} M. Lavagna,
                \textit{Phys. Rev. B} \textbf{1990}, \textit{41}, 142.
                        
\bibitem{Li94} T. Li and P. B\'enard, 
               \textit{Phys. Rev. B} \textbf{1994}, \textit{50}, 17837.  
   
\bibitem{Jol91} Th. Jolic{\oe}ur and J. C. Le Guillou,
                \textit{Phys. Rev. B} \textbf{1991}, \textit{44}, 2403.
    
                           
\bibitem{Kot92} Y. Bang, C. Castellani, M. Grilli, G. Kotliar, R. Raimondi,
                and Z. Wang,
                \textit{Int. J. of Mod. Phys. B} \textbf{1992}, \textit{6}, 531;
                 Proceedings of the Adriatico Research Conference and
                 Miniworkshop \textit{Strongly Correlated Electrons Systems
                 III}, eds. Yu Lu, G. Baskaran, A. E. Ruckenstein, E.
                 Tossati (World Scientific Publishing Co., Singapore, 1992).   
                            
\bibitem{Dao17}  V.~H.~Dao and R.~Fr\'esard, 
                 \textit{Phys. Rev. B} \textbf{2017}, \textit{95}, 165127. 
                 
\bibitem{Dao18}  V.~H.~Dao and R.~Fr\'esard,
                 \textit{Acta Phys. Pol. A} \textbf{2018}, \textit{133}, 336. 
  
                 
\bibitem{RFTK12} R.~Fr\'esard and T.~Kopp, 
                 \textit{Ann. Phys. (Berlin)} \textbf{2012}, \textit{524}, 175.
                                            
\bibitem{Bickers86} N. E. Bickers, Ph.D. Thesis, Cornell University, 1986.

\bibitem{Bickers87} N. E. Bickers,  
              \textit{Rev. Mod. Phys.} \textbf{1987}, \textit{59}, 845.
                    
\bibitem{negele}  J.~W. Negele and H.~Orland, 
 \textit{Quantum Many-Particle Systems} (Westview Press, Boulder, Colorado, 1998).
   

\end{thebibliography}
\end{document}